\documentclass[aps,prc,twocolumn,showpacs,superscriptaddress,footinbib]{revtex4-2}
\usepackage[normalem]{ulem}

\usepackage{graphicx}
\usepackage{epstopdf}
\usepackage{calrsfs}
\usepackage{dcolumn}
\usepackage{bm}
\usepackage{enumerate}
\usepackage{amsmath}

\usepackage[normalem]{ulem}  
\usepackage[colorlinks]{hyperref}

\usepackage{amssymb}
\usepackage[mathscr]{eucal}
\usepackage{lipsum} 
\usepackage{ulem}

\begin{document}


\title{Complete fusion of $^6$Li with $^{28}$Si and $^{64}$Ni nuclei in the framework of the continuum discretized coupled channel method}

\author{S.R. Souza}
\email{Corresponding author: srsouza@if.ufrj.br}
\affiliation{Instituto de F\'\i sica, Universidade Federal do Rio de Janeiro Cidade Universit\'aria, 
\\Caixa Postal 68528, 21941-972 Rio de Janeiro-RJ, Brazil}
\affiliation{Departamento de F\'\i sica, ICEx, Universidade Federal Fluminense,
\\R. Desembargador Ellis Hermydio Figueira, Aterrado, 27213-145 Volta Redonda - RJ, Brazil}
\author{L.F. Canto}
\affiliation{Instituto de F\'\i sica, Universidade Federal do Rio de Janeiro Cidade Universit\'aria, 
\\Caixa Postal 68528, 21941-972 Rio de Janeiro-RJ, Brazil}
\author{R. Donangelo}
\affiliation{Instituto de F\'\i sica, Universidade Federal do Rio de Janeiro Cidade Universit\'aria, 
\\Caixa Postal 68528, 21941-972 Rio de Janeiro-RJ, Brazil}
\affiliation{Instituto de F\'\i sica, Facultad de Ingenier\'\i a, Universidad de la Rep\'ublica, 
Julio Herrera y Reissig 565, 11.300 Montevideo, Uruguay}

\date{\today}

\begin{abstract}
The complete fusion of $^6$Li with $^{28}$Si and  $^{64}$Ni nuclei, ranging from subbarrier energies up to values well above the Coulomb barrier, is studied using the  continuum discretized coupled channels method.
We investigate the sensitivity of the results to the largest energy used in the discretization of the continuum, including closed channels.
Our results reveal that, as long as the states in the continuum are not limited to too small energies, the specific upper bound adopted is not relevant as the complete fusion cross-section is fairly insensitive to it above a threshold.
Convergence with respect to this parameter is rapidly reached, so that closed channels play a role only at low collision energies.
A good agreement with the available experimental data is obtained.
\end{abstract}

\maketitle


\begin{section}{Introduction}
\label{sect:introduction}

%

%
Collisions involving weakly bound projectiles constitute an important tool for 
studying the effects of their dissociation on different reaction mechanisms 
\cite{CGD15,HOM22,Mor18}.
Such nuclei may be excited to unbound states during the collision so that the 
scenario of two 
\cite{CGD15,HOM22,Mor18,Tho88,YNI82,
DBD10} or three \cite{Mor18,DDC15,DDC15a,MHY04,
RAG08,RAG09} fragments interacting with the target must be considered.\\

The low binding energy of the projectile affects fusion reactions in two ways. First, it leads to a reduction of the Coulomb barrier, arising from the long 
tail of the projectile's density. This is a static effect that enhances fusion at all collision energies. On the other hand, the low breakup threshold has a 
strong influence on the reaction dynamics, giving rise to new reaction channels. 
One possible outcome is the dissociation of the projectile 
without the absorption of any of the fragments by the target \cite{CGD15,Mor18,DBD10,TNT01,Huu15,CGY22}.
Alternatively, the partial fusion of the projectile, {\it i.e.}, the absorption of some, but not all of the fragments, leads to the so called incomplete fusion 
(ICF). There are different ICF processes, characterized by the particular fragment that is captured by the target.  There is also the process where all 
the breakup fragments fuse with the target, which is called sequential complete fusion (SCF). Another possible scenario is the direct complete fusion 
(DCF), in which the entire projectile fuses with the target without previously dissociating. Although these mechanisms may be treated within the 
framework of theoretical approaches \cite{CGD15,HOM22}, the SCF cannot be distinguished experimentally from the DCF as both lead to the same 
compound nucleus, and their sum is named complete fusion (CF).
The excitation of these unbound states reduces DCF, which is usually the leading fusion mechanism. This leads to a suppression of CF
at all collision energies. Then, the net effect of the low binding energy on the CF
cross section is the result of these competing effects. Comparisons of theoretical and experimental CF cross sections with predictions of 
barrier penetration models that disregard the cluster structure of the projectile indicate suppression above the Coulomb barrier and enhancement
at sub-barrier energies~\cite{CGD15,HOM22,Mor18}. \\

Different theoretical approaches have been developed to study the fusion processes taking place in collisions of weakly bound projectiles (see
the review articles~\cite{CGD15,HOM22,Mor18}, and references therein). The most realistic ones are based on the Continuum Discretized Coupled 
Channel method (CDCC), to deal with coupled channel equations involving the 
continuum~\cite{SYK83,SYK86,AIK87,ThN09,MHO04,RAG08,RAG09,DDC15,MHO04,RAG08,RAG09,DDC15,YIK86,GoM17}. Presently,
a few computer codes to  implement this method are available~\cite{Tho88,HRK99,Des16}. 
 However, the CDCC alone does not give individual cross sections for the CF or ICF processes. For this purpose, complementary assumptions must be made. 
It is also worth mentioning the indirect determination of the CF cross section for the $^{6,7}$Li + $^{209}$Bi systems~\cite{LeM19}, which was extracted 
from the total reaction cross section, by subtracting its inelastic, elastic breakup, and inclusive nonelastic breakup components. \\

The first CDCC based calculations of CF and ICF cross sections were made by Hagino {\it et al.}~\cite{HVD00} and by Diaz-Torres and Thompson~\cite{DiT02},
for the $^{11}{\rm Be}$ + $^{208}$Pb system.  $^{11}{\rm Be}$ is  a weakly bound nucleus, that breaks up into a $^{10}$Be cluster and one neutron.
 These works adopted a short-range imaginary potential depending exclusively on the projectile-target distance. Their basic assumption was
 that absorption in the bound channels contributed to CF, whereas absorption in unbound channels corresponded to ICF of the $^{10}$Be cluster. 
 Since this cluster is much heavier than the neutron, in an unbound channel, it must be close to the projectile, while the neutron is far away. 
 Then, it is reasonable to assume that
absorption by the imaginary potential corresponds to ICF. However, this approach has important limitations. First, it cannot be used for projectiles
like  $^{6,7}$Li, which break up into fragments with comparable masses. Besides, this method cannot determine individual cross section for the two
ICF processes which, in principle, could be determined experimentally.\\

For the calculation of individual ICF cross sections, the imaginary potential must be a sum of terms, each one accounting for the absorption of 
a fragment of the projectile. Diaz-Torres, Thompson and Beck~\cite{DTB03}, and Descouvemont {\it et al.}~\cite{DDC15} performed 
CDCC calculations using an imaginary potential of this kind. However, they only evaluated the TF cross section, which was identified with the total 
absorption of the imaginary potential. \\

Recently, Rangel {\it et al.}~\cite{RCL20} developed a CDCC based method to evaluate the CF and ICF cross sections, henceforth referred to as the CF-ICF model. 
Their CDCC calculations were similar to those of Ref.~\cite{DTB03}, but they made additional assumptions to relate the observable cross sections to 
quantities that could be obtained from the CDCC calculations. This method was applied to collisions of several weakly bound projectiles on heavy 
targets~\cite{RCL20,CRF20,LFR22,FRL23}. In the CF-ICF model, the CF and ICF cross sections are expressed in terms of radial integrals of the imaginary
potentials. These integrals, require scattering wave functions in bound and unbound channels, evaluated with great accuracy in the internal region of the
Coulomb barrier. In Refs.~\cite{RCL20,RCL20,CRF20,LFR22,FRL23}, the wave functions were obtained by FRESCO, which discritize the continuum
by the {\it bin method}~\cite{Tho88}. This code is optimized to give accurate
elastic and breakup cross sections. These cross sections are expressed in terms of the S-matrix, which is extracted from the asymptotic forms of the
wave functions. Thus, it is not particularly concerned with wave functions in the inner region of the barrier. For this reason, the convergence of the CF and ICF
cross sections of Refs.~\cite{RCL20,RCL20,CRF20,LFR22,FRL23} demanded great care, mainly at energies well below the Coulomb barrier.\\

In the present work, we discuss a new implementation of the CF-ICF model, where the CDCC equations are solved by the  R-matrix formalism of Ref.~\cite{DeB10}, and the CF and ICF cross sections are calculated by the expressions in the appendix of Ref.~\cite{CRF20}. Then, we evaluate
the CF cross section for collisions of $^6$Li with targets lighter than the ones studied in previous applications of the CF-ICF model.
The remainder of this manuscript is organized as follows. The main features of the CDCC treatment are reviewed in Sect.\ \ref{sect:model}, 
whereas the main results are presented in Sect.\ \ref{sect:results}. A brief summary of our main findings is made in Sect.\ \ref{sect:conclusions}.

\end{section}

 
\begin{section}{Theoretical framework}
\label{sect:model}


We consider the collision of a projectile that dissociates into two fragments, $c_1$ and $c_2$. In the case of $^6$Li, $c_1$ is the deuteron and $c_2$ is the alpha particle.
The real part of the projectile-target interaction has the form,
\begin{equation}
V(\vec{R},\vec{r}) = V_1(r_1) + V_2(r_2),
\label{V1-V2}
\end{equation}
where $r_i$ is the distance between $c_i$ and the target, and $V_i(r_i)$ is the corresponding potential. Throughout the present work, these interactions are
given by the Aky\"uz-Winther potential~\cite{BrW04,AkW81}.  The vectors $\vec{r}_1$ and $\vec{r}_2$ are related to the vector between the projectile and
the target, $\vec{R}$, and the vector joining the two fragments,  $\vec{r}$, by the relations\\
\[
\vec{r}_1 = \vec{R} + \frac{2}{3}\ \vec{r},\qquad \vec{r}_2 = \vec{R} - \frac{1}{3}\ \vec{r} .
\]

For the imaginary part of the potential, we follow Ref.~\cite{CRF20}, adopting different forms for the bound and unbound channels. The details can be
found in  Ref.~\cite{CRF20}.\\

In the CDCC  method \cite{CGD15,Tho88}, the scattering wavefunction $\Psi^{(+)}(\vec{k}_0,\vec{R})$ describing the colliding nuclei, whose incident 
relative momentum  is $\hbar\vec{k}_0$, is expanded on eigensates $|\varphi_{\alpha,i})$ of the intrinsic projectile's hamiltonian and scattering states $\psi_{\alpha,i}^{(+)}(\vec{k}_0,\vec{R})$, projected on channel $\left\{\alpha,i\right\}$:

\begin{equation}
\Psi^{(+)}(\vec{k}_0,\vec{R})=\sum_{\alpha,i}\psi^{(+)}_{\alpha,i}(\vec{k}_0,\vec{R})|\varphi_{\alpha,i})\;,
\label{eq:Psi}
\end{equation}
where $\alpha$ denotes the set of quantum numbers associated with the channel and $i$ represents the index of an energy in the set of the discrete values used for each $\alpha$.
From this expansion, one may write the set of equations which couple the different channels considered in the calculation \cite{DeB10}, with the same total angular momentum $J$ and parity $\pi$:

\begin{gather}
\left[-\frac{\hbar^2}{2\mu}\left(\frac{d^2}{dR^2}-\frac{L(L+1)}{R^2}\right)+(\varepsilon_i^{jl}-\varepsilon_0)-E_{\rm CM}\right]\phi^{J\pi}_{\alpha,i}(R)\nonumber\\ 
= \sum_{\alpha',i'}V_{\alpha i,\alpha'i'}^{J\pi}(R)\phi^{J\pi}_{\alpha',i'}(R)\;.
\label{eq:CCeqs}
\end{gather}

\noindent
Above, $\mu$ represents the reduced mass of the projectile and target nuclei,  $L$ denotes the quantum number associated with their relative orbital angular momentum, and $\phi^{J\pi}_{\alpha,i}(R)$ symbolizes the radial component of $\psi^{(+)}_{\alpha,i}(\vec{k}_0,\vec{R})$.
The center of mass kinetic energy of the colliding partners and the projectile's intrinsic energy in the entrance channel are respectively denoted by  $E_{\rm CM}$ and $\varepsilon_0$.
The $i$-th eigenvalue of the intrinsic projectile's hamiltonian, associated with a relative orbital ($l$) and total ($j$) angular momenta quantum numbers, is represented by $\varepsilon_i^{jl}$.
After performing the angular integrations over the projectile-target and fragment-core coordinates on the multipole expansion of the potential acting between the target and the clusters,  the coupling between the different channels $V^{J\pi}_{\alpha i,\alpha'i'}(R)$, is given by \cite{DBD10}:

\begin{gather}
V^{J\pi}_{\alpha i,\alpha'i'}(R)=(-1)^{J-s}\,i^{l'+L'-l-L}\,\hat \j \hat\j'\hat l\hat l'\hat L\hat L' \nonumber \\
\times \sum_\lambda(-1)^\lambda
\begin{pmatrix}
l & l ' & \lambda\\
0 & 0 & 0
\end{pmatrix}
\begin{pmatrix}
L & L ' & \lambda\\
0 & 0 & 0
\end{pmatrix}
\begin{Bmatrix}
j & L & J\\
L' & j ' & \lambda
\end{Bmatrix}\nonumber\\
\times
\begin{Bmatrix}
j & l & s\\
l ' & j ' & \lambda
\end{Bmatrix}
\int_0^\infty\,dr\,\tilde\varphi^{jl}_i(r)V_\lambda(r,R)\tilde\varphi^{j'l'}_{i'}(r)\;.
\label{eq:VJp}
\end{gather}
 
\noindent
In this expression, $V_\lambda(r,R)$ corresponds to the coefficient of the multipole expansion \cite{CRF20} and $\tilde\varphi_i^{jl}(r)$ is the radial part of the projectile internal wavefunction.
The quantum numbers $l$, $j$, and $L$ are associated with the index $\alpha=\left\{l,j,L\right\}$.
The spin of the projectile's fragment is denoted by $s$ and we consider spinless cores.
In the present case, $s=1$, as we focus on the dissociation of the $^6$Li into a deuteron and a ground state $^4$He nucleus.
We adopt the usual notation $\hat x=\sqrt{2 x+1}$, besides the representation $()$ and $\left\{ \right\}$ for the Wigner 3J and 6J symbols, respectively.\\

In collisions of $^6$Li with tightly bound targets, which will be considered in the present work, the off-diagonal matrix elements of Eq.~(\ref{eq:VJp}) couple the elastic channel (the only bound channel of the system) to the continuum. On the other hand, the 
diagonal matrix element of the potential, 
\begin{equation}
V_{00}(R) = \left( \varphi_0\,\left| \,V_1\,+\,V_2\,\right| \varphi_0\right),
\label{V00}
\end{equation}
where $\varphi_0$ stands for the ground state of $^6$Li, plays the role of the real part of the optical potential in the elastic channel. 
The barrier of this potential is denoted by $V_{00}^B$. \\

It is interesting to compare $V_{00}^B$ to the barrier of the $^6$Li-target AW
potential when the cluster structure of the projectile is neglected. This potential and its Coulomb barrier are denoted by $V_{PT}(R)$ 
and $V_{PT}^B$, respectively. 
\begin{table}
\centering
\caption{Barriers of the $V_{PT}(R)$ and $V_{00}(R)$ potentials (in MeV), for the systems studied in this work.}
\vspace{0.5cm}
\begin{tabular}{lcccc}
\hline 
System                        \qquad\qquad\qquad \  \    &  $V_{PT}^B$ \qquad\qquad \   &\qquad\  $V_{00}^B$   \\ 
\hline
$^6$Li + $^{28}$Si      \qquad\qquad\qquad\ \  \   &    7.0            \qquad\qquad\     &  \qquad\ 6.4               \\
$^6$Li + $^{64}$Ni     \qquad\qquad\qquad\ \ \    &     12.4         \qquad\qquad\     & \qquad\ 11.6                \\
%
\hline
\end{tabular}
\label{barriers}
\end{table}
Table~\ref{barriers} shows the barriers of the two potentials, for the systems studied in the present work. One
concludes that the reductions for the $^6$Li + $^{28}$Si and $^6$Li + $^{64}$Ni systems are of
0.6 and 0.8 MeV, respectively.


\begin{subsection}{Intrinsic projectile states}
\label{subsect:projstat}


To obtain the intrinsic projectile states, we follow Refs.\ \cite{DBD10,Bay06} and expand $\tilde\varphi^{jl}_i(r)$ on a Lagrange-Laguerre basis:

\begin{equation}
\tilde\varphi^{jl}_i(r)=\frac{1}{h^{1/2}}\sum_{k=1}^N c_k^{jli}f_k(r/h)\;,
\label{eq:phiexp}
\end{equation}

\noindent

where

\begin{equation}
f_k(x)=(-1)^j x_k^{1/2}\left(\frac{x}{x_k}\right)\frac{L_N(x)}{x-x_k}\exp[-x/2]\;,
\label{eq:scaledBasis}
\end{equation}

\noindent
$L_N(x)$ is the Laguerre polynomial of order $N$, $x_k$ its $k$-th root and $h$ is a scaling parameter.
One should note that the use of this basis leads to square integrable wavefunctions due to the exponential factor.
To obtain the bound and continuum internal pseudo states of the projectile, the expansion given by Eq.\ (\ref{eq:phiexp}) is inserted into the eigenvalue equation associated with the internal projectile hamiltonian. This leads to \cite{Bay06}:

\begin{equation}
\sum_{k'=1}^N\left[T_{k,k'}^l+v_{jl}(x_k h)\delta_{kk'}\right]c_{k'}^{jli}= c_k^{jli}\varepsilon_i^{jl}\;,
\label{eq:pseudoStates}
\end{equation}

\noindent
where $\delta_{kk'}$ corresponds to the Kornecker delta function, $T^l_{k,k}$ denotes the matrix elements of the kinetic energy operator acting on the relative coordinates of the projectile fragments, and $v_{jl}(x_k h)$ is the total potential between them at separation $r=x_kh$.
Explicit expressions for $T^l_{kk'}$ may be found in Ref.\  \cite{Bay06}.
The solution of this system of equations leads to the eigenvalues  $\left\{\varepsilon_1^{jl},\cdots,\varepsilon_N^{jl}\right\}$ and the eigenvectors
$\left\{(c^{jl,1}_1,c^{jl,1}_2,\cdots,c^{jl,1}_N),\cdots,(c^{jl,N}_1,c^{jl,N}_2,\cdots,c^{jl,N}_N)\right\}$.
The latter are needed to build $\tilde\varphi^{jl}_i(r)$ through Eq.\ (\ref{eq:phiexp}).
Therefore, the number of states considered in the calculation depends on $N$ and $h$, which are parameters of the calculation.
However, the final results must be stable for relative small variations on specific choices of these parameters.

The potential energy above is split into two terms.
The Coulomb one reads:

\begin{equation}
v_{\rm Coul}(r)=
\begin{cases}
\frac{Z_{c_1} Z_{c_2} e^2}{r}\;,\;\; r\ge R_c\\
\frac{Z_{c_1} Z_{c_2} e^2}{2R_c}\left[3-\frac{r^2}{R_C^2}\right]\;,\;\; r \le R_C\;,
\end{cases}
\label{eq:Coulomb}
\end{equation}

\noindent
where $R_C=r_{0C}(A_{c_1}^{1/3}+A_{c_2}^{1/3})$ and $r_{0C}=1.3$ fm.
Above, $c_i$ represents the {\it i}-th cluster and $A_{c_i}$ ($Z_{c_i}$) is its mass (atomic) number.
As in Ref.\ \cite{KCD18},  the nuclear interaction between the clusters is approximated by:

\begin{equation}
v_{\rm nuc}(r)=-V_0 F(r)+A_{SO}\left(\frac{\hbar}{m_\pi c}\right)^2\frac{1}{r}\frac{dF(r)}{dr}\;\vec{l}\cdot\vec{s}\;,
\label{eq:vnuc}
\end{equation}

\noindent
and

\begin{equation}
F(r)=\frac{1}{1+\exp\left[(r-R_0)/a\right]}\;,
\label{eq:FSO}
\end{equation}

\noindent
where $m_\pi$ is the pion mass.
Since in this work we consider the $^6$Li projectile, we adopt the following parameters for the nuclear interaction between the clusters \cite{LFR22}:
$R_0=r_0(A_{c_1}^{1/3}+A_{c_2}^{1/3})$, $r_0=1.0046$ fm, and $a=0.7$ fm in all cases.
We employ $V_0=78.38$ MeV and $A_{SO}=0$ for $j=1$ and $l=0$, otherwise $V_0=80.0$ MeV and $A_{SO}=2.5$ MeV.
This choice allows one to reproduce the experimental dissociation energy ($1.47$ MeV) of the $^6$Li into the deuteron and the ground state $^4$He nucleus, with $J^\pi=1^+$ ($l=0$), as well as its resonances \cite{LFR22,DTB03}.
The energy of the pseudo states obtained with these parameters, using $N=35$ and $h=0.5$, are exhibited in Fig.\ \ref{fig:levels} for selected values of $j$ and $l$.
The results show that the density of states is much smaller at energies above 5 MeV than below it.
Therefore, one should expect important contributions to the continuum-continuum coupling from such low lying energy states.

\begin{figure}[tbh]
\includegraphics[width=8.5cm,angle=0]{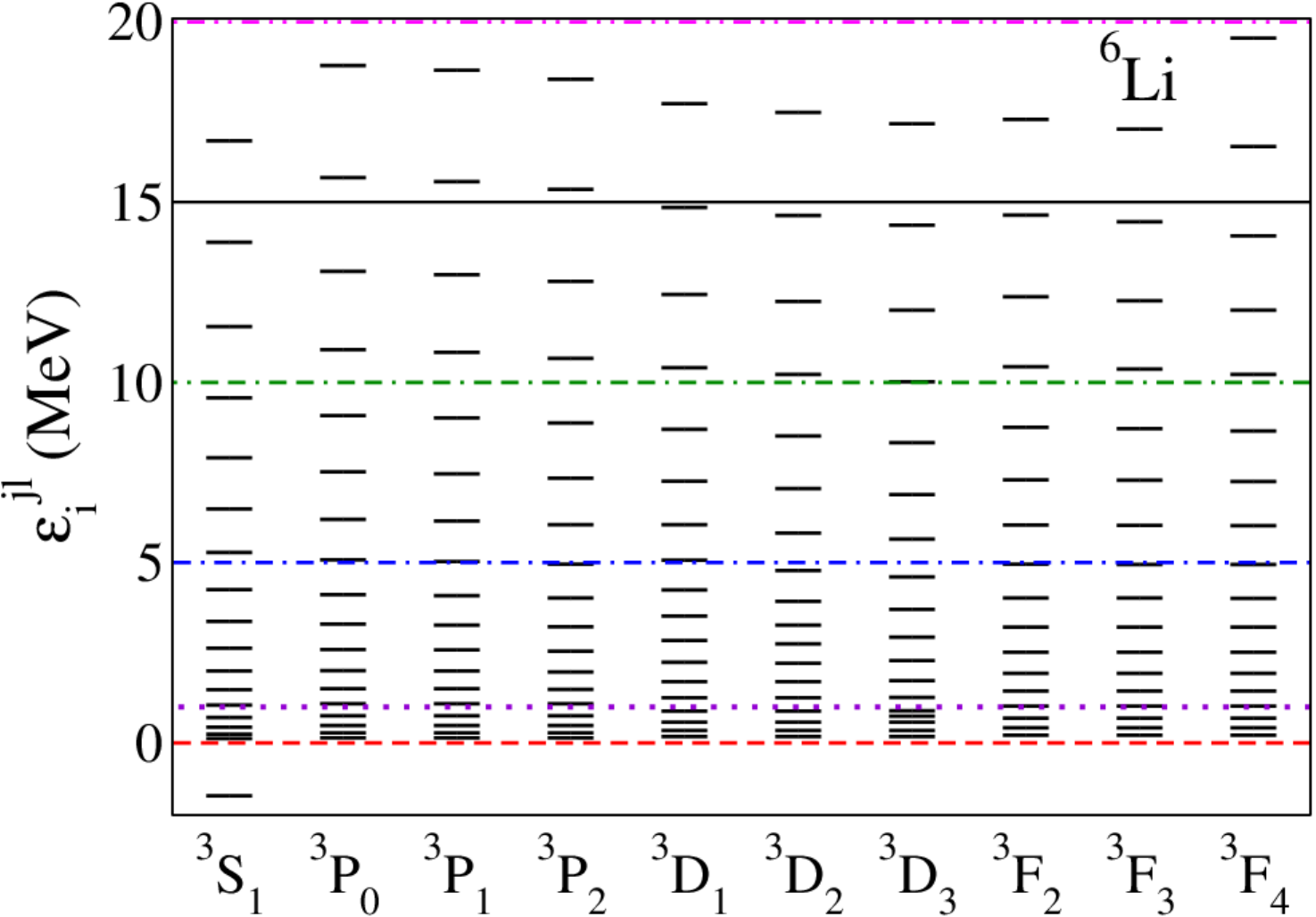}
\caption{\label{fig:levels} (Color online) Energy levels of pseudo states, for different $j$ and $l$, describing the two cluster states of $^6$Li nucleus for $N=35$ and $h=0.5$.
For details, see the text.}
\end{figure}

\end{subsection}


\begin{subsection}{Channel wavefunctions and the fusion cross-section}
\label{sect:cs}


To solve Eqs.\ (\ref{eq:CCeqs}), the potentials needed to calculate the multipole coefficients $V_\lambda(r,R)$ must be specified.
Besides the Coulomb one, whose parameterization is similar to that given by Eq.\ (\ref{eq:Coulomb}), one adopts, as in Ref.\ \cite{KCD18},
the Aky\"uz-Winther potential (AW) \cite{BrW04,AkW81} for the nuclear  interactions between the fragments and the target.\\

Different techniques have been employed to solve the system of coupled equations (\ref{eq:CCeqs}) \cite{Tho88,ARM77,CGY22,MKO03,YNI82,DeB10,RMP80}.
Owing to the robustness of its numerical solutions, we follow Refs.\ \cite{DeB10,DBD10,Des16} where the application of the R-matrix theory to this problem is carefully described.
A difficulty of this treatment is that, for each value of $J^\pi$ and $L$, it requiries the inversion of a large matrix, whose size depends on the number of the radial mesh points adopted in the calculations, multiplied by the number of channels included.
For a fine discretization of the radial mesh and a large number of channels, very large matrices are rapidly obtained and their inversion may be prohibitively time consuming.
To mitigate this difficulty, the space is divided into ${\cal N}_s$ spherical shells delimited by $a_{k-1} \le R \le a_k$, $1 \le k\le {\cal N}_s$, $a_0=0$, $a_{{\cal N}_s}=R_{\rm max}$.
The latter denotes the largest radial distance on the mesh.
Within the {\it k}-th shell, $N_k$ points are used in the discretization.
In this way, one needs to invert ${\cal N}_s$ matrices of much smaller sizes, which considerably speeds up the numerical implementation.
As is detailed described in  Ref.\ \cite{Des16},  $\phi^{J\pi}_{\alpha,i}(R)$ is expanded on a Gauss-Lagrange basis within each shell:

\begin{equation}
\phi^{J\pi}_{\alpha ,i}(R)=\sum_{n=1}^{N_k} c_n^{J\pi\alpha,i}(k)g^{(k)}_n(R),\;\;\; a_{k-1} \le R\le a_k.
\label{eq:phiRexp}
\end{equation}

\noindent
In the above expression, the base function $g^{(k)}_n(R)$ is:

\begin{eqnarray}
g^{(k)}_n(R)&=&(-1)^{N_k+k}\sqrt{\Delta_k x_n^{(k)}(1-x_n^{(k)})}\nonumber\\
&\times&\left(\frac{R}{a_{k+1}x_n^{(k)}}\right)^{\delta_{k,1}}\frac{P_{N_k}\left(\frac{2R-a_k-a_{k-1}}{\Delta_k}\right)}{R-\Delta_k x_n^{(k)}-a_k},
\label{eq:gbasis}
\end{eqnarray}

\noindent
where $\Delta_k=a_k-a_{k-1}$, $P_{N_k}(x)$ is the Legendre polynomial of order $N_k$, and $x_n^{(k)}$ is the {\it n}-th root of $P_{N_k}(2x_n^{(k)}-1)=0$.
This basis leads to efficient evaluation of the matrix elements \cite{DeB10} entering the calculations of the multipole expansion $V_\lambda(r,R)$ \cite{CRF20}.

The numerical implementation of the R-matrix method with Lagrange meshes is carefully described in Refs.\ \cite{Des16,DeB10}.
Since our implementation follows these references, we do not furnish further details.
One should note that the quantities ${\cal N}_s$, $\left\{a_1,\cdots,a_{{\cal N}_s}\right\}$, $\left\{N_1,\cdots,N_{{\cal N}_s}\right\}$, and $R_{\rm max}$ are parameters of the calculations, whose influence on the final results must be checked in actual applications.\\

For the calculation of the CF cross section, we use the version of the CF-ICF model presented in Ref.~\cite{LFR22}. The cross section is evaluated in three steps.
\begin{enumerate}

\item
Using the expressions in the appendix of Ref.~\cite{CRF20}, we evaluate the DCF probability for each angular momentum, $P^{DCF}_J$. We also 
evaluate individual fusion probabilities for each fragment, denoted by $P^{(1)}_J$ and $P^{(2)}_J$. Note that $P^{(1)}_J$ and $P^{(2)}_J$ are inclusive 
probabilities. That is, the probability of fusion of one of the fragments, whatever happens to the other. 

\item
Then, using the above results, we obtain the observable CF probability, $P^{CF}_J$, by adopting the intuitive assumption of Ref.~\cite{LFR22}, 
\begin{equation}
P_J^{SCF} = P^{(1)}_J P^{(2)}_J,
\label{eq:ptf}
\end{equation}
and thus
\[ 
P^{CF}_J=P^{DCF}_J+P^{(1)}_J P^{(2)}_J.
\]
\item
Finally, the CF cross section is obtained by carrying out the angular momentum sum
\begin{equation}
\sigma_{CF}=\frac{\pi}{k^2_0}\sum_J (2J+1)P^{CF}_J\;.
\label{eq:sigCF}
\end{equation}
\end{enumerate}

 We emphasize that we adopt the prescription of Ref.~\cite{CRF20} for the imaginary potential. For bound channels, it depends exclusively on the 
 projectile-target distance, and is denoted by $W_{\rm PT}(R)$. For unbound channels, it is a sum of two terms, each one representing the absorption 
 of one of the fragments by the target. It is denoted by $W_1(r_1) +   W_2(r_2)$. The details of these potentials can be found in Ref.~\cite{CRF20}.\\

Finally, as in Ref.\ \cite{LFR22}, we introduce the spectroscopic amplitude $\alpha$ in the coupling between bound and unbound states of the $^6$Li 
clusters, in order to take into account the fact that the d + $^4$He cluster structure is an approximation to the actual ground state of the nucleus.
We adopt the same value $\alpha=0.7$ used in that study.

\end{subsection}

\end{section}

\begin{section}{Results}
\label{sect:results}
Before discussing the main results of the present investigation, we report the tests made on their sensitivity to the parameters of the calculations.
For simplicity, we use shells such that $a_{k+1}-a_k= 5.0$ fm and $N_k=20$, $1\le k\le {\cal N}_s$, but different spacing and number of points may be adopted for
each one.
We have found that finer discretizations do not affect the results presented in this work.
Our tests also suggest that $R_{\rm max} = 150$ fm (${\cal N}_s=30$) is sufficient for the results presented below.

Since the coupling with states in the continuum plays a relevant role in the fusion process involving weakly bound nuclei \cite{DTB03,CGY22,LFR22,CRF20,AIK87,HOM22}, the inclusion of such states must be carefully considered.
Therefore, we examined the sensitivity of our results to the discretization of the continuum using different values of $(j,l)$.
For both systems considered in the present study, we find that, at energies below the Coulomb barrier, varying the largest value of $l$ ($l_{\rm max}$) from 3 to 4 changes the CF cross-section approximately $10\%$.
The differences decrease very rapidly as the collision energy increases, being negligible at a few MeV above the barrier.
Thus, the use of $l_{\rm max}=3$ is a good compromise between computational efficiency and precision.
One should note that, if energies in the continuum up to $\varepsilon_{\rm max}=20$ MeV are allowed, this leads to a system of more than 600 coupled equations for $l_{\rm max}=4$.
There are, in this way, ${\cal N}_s$ complex matrices of size $600N_k\times 600 N_k$ to be inverted for each $(J^\pi,L)$.\\

\begin{figure}
\includegraphics[width=8.5cm,angle=0]{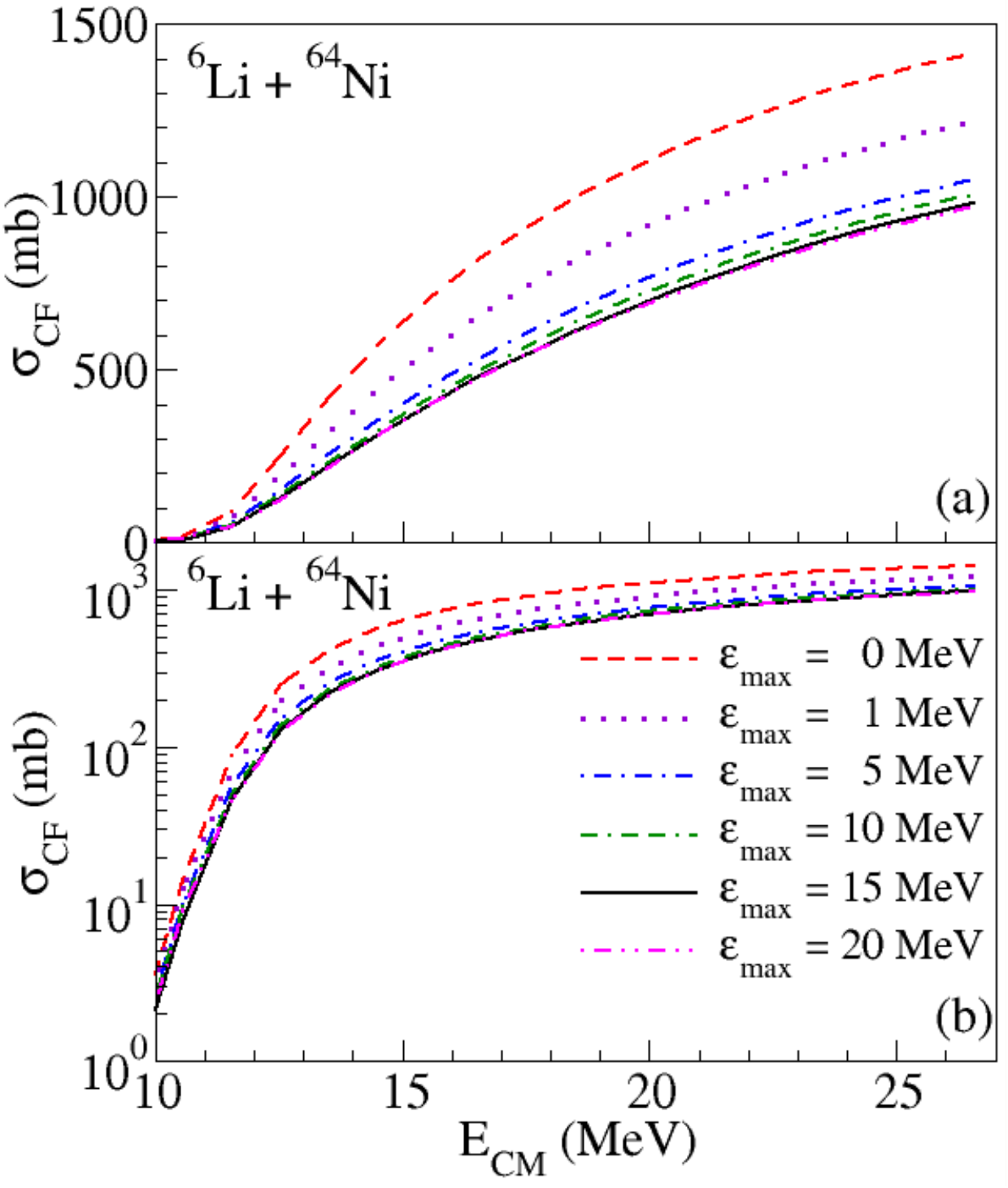}
\caption{\label{fig:sigCFEc} (Color online) Complete fusion cross-section for different values of $\varepsilon_{\rm max}$, for collisions between the $^6$Li and the $^{64}$Ni nuclei at different values of the center of mass energy.
Panel (a) displays the results in linear scale whereas the semi-log scale is used in panel (b).
For details, see the text.}
\end{figure}
To investigate the influence of  $\varepsilon_{\rm max}$ on the fusion cross-section, we show, in Fig.\ \ref{fig:sigCFEc}, the CF cross-sections for the 
$^6$Li + $^{64}$Ni system as a function of $E_{\rm CM}$ obtained using different values of $\varepsilon_{\rm max}$.  We consider 
collision energies ranging from 10 to 27 MeV. The lowest energy corresponds to $\sim$ 2 MeV below the Coulomb barrier of the folded potential, $V_{00}(R)$,
given in Table~\ref{barriers}. 
The results reveal that the CF cross sections is 
fairly sensitive to this parameter. This is because the coupling with states in the continuum redistributes flux among the channels, removing contributions 
from the bound to the unbound ones. The latter leads to the incomplete fusion of one of the fragments \cite{LFR22}, at the cost of lowering the DCF and, 
consequently, the CF if the contribution due to SCF does not tip the balance in favor of CF. This is the case for the systems studied in this work.
Nevertheless, the  differences between the results for $\varepsilon_{\rm max}=10$ and $\varepsilon_{\rm max}=20$ MeV are typically of the order of
5\%, throughout the energy interval of the calculations.
Convergence is practically reached for 10 MeV $ \le \varepsilon_{\rm max}\le $ 20 MeV. Our findings are in agreement with those of 
Refs.~\cite{DTB03,LFR22} that obtained converged results using similar upper bounds.
The fact that the largest changes occur for $\varepsilon_{\rm max} \le 5$ MeV, even for $E_{\rm CM} \gg 5$ MeV, does not mean that closed channels 
do not play a relevant role in the coupling among the channels. Indeed, we have checked that, for $\varepsilon_{\rm max}=15$ MeV, the CF cross-section
at $V_{00}^B-2$ MeV changes by about 20\% if all closed channels are excluded from the calculation.
They correspond to approximately 24\% of the total number of channels. 
Note that setting $\varepsilon_{\rm max} = 0$ is equivalent to switching off all couplings with the continuum. Then, the set of coupled 
equations reduces to a single equation for the elastic channel, with the potential $V_{00}(R)$. In this way, the dashed line in Fig.~\ref{fig:sigCFEc} 
is the cross section of a one-channel calculation with this potential, which is virtually the same as the cross section of the barrier penetration 
model (BPM) for the same potential~\cite{CDH18}. This cross section contains the static effects of the barrier lowering, resulting from the low breakup threshold
of $^6$Li. Thus, comparing it to the CF cross section one finds that the latter is suppressed at all collision energies. This is a consequence of
the breakup couplings on complete fusion. 

Contrary to the conclusions of Ref.\ \cite{CPG23}, our results suggest that closed channels affect the fusion cross-section only at low collision energies as convergence with respect to $\varepsilon_{\rm max}$ is reached at relatively small values, compared with those obtained in that work.
By contrast, our results reveal that the coupling with low lying states in the continuum are more relevant in this context since the CF cross-section is very sensitive to states with energy $\varepsilon_{\rm max} \lesssim 5$ MeV.
Therefore, as long as one is interested in the fusion process, we find that there is no need to include very high energy continuum states in CDCC calculations.

\begin{figure}
\includegraphics[width= 8.7cm,angle=0]{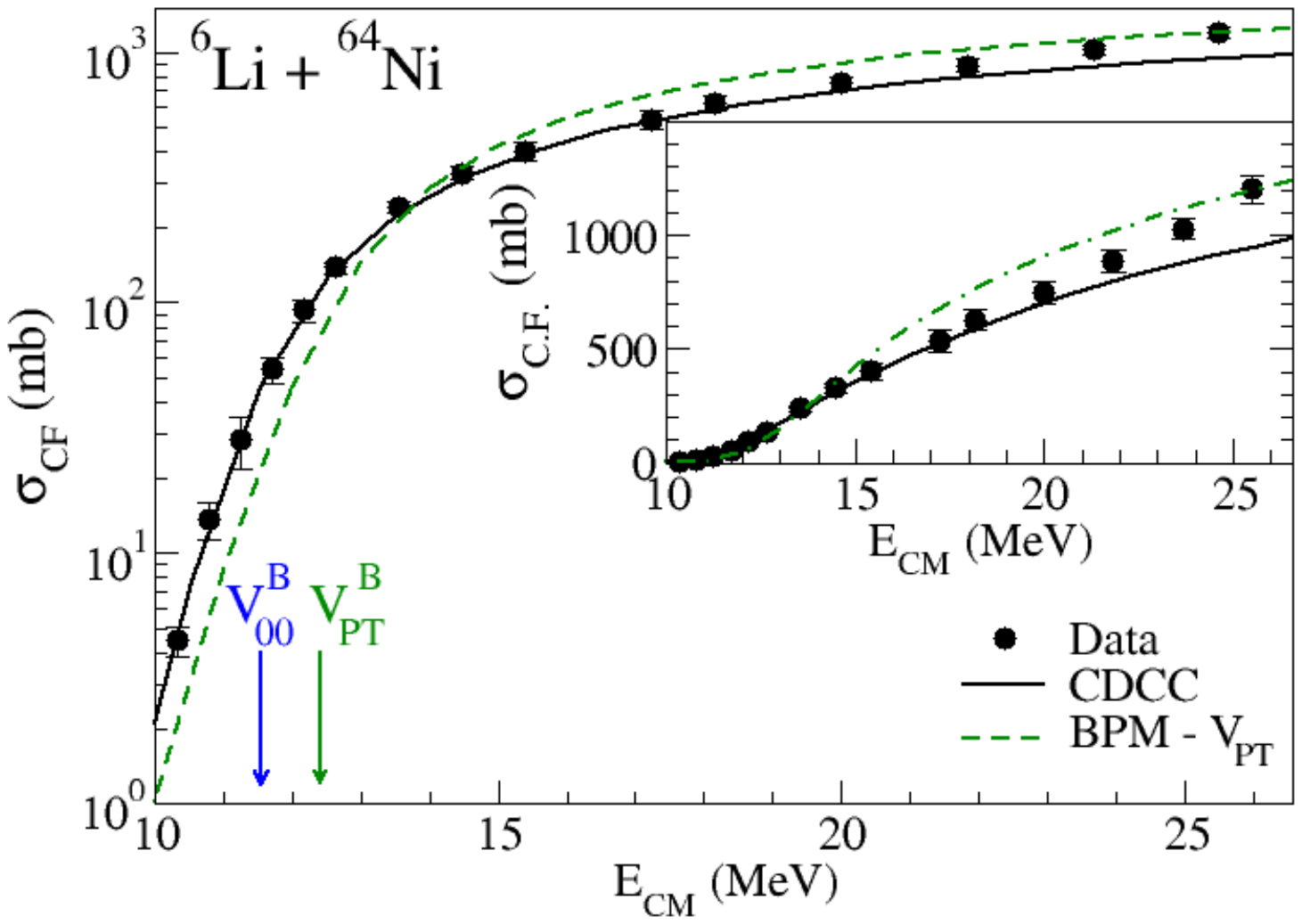}
\caption{\label{fig:sigCFNi} (Color online) Complete Fusion cross-section as a function of the center of mass energy of the colliding nuclei.
The data are from Ref.\  \cite{MRR14}.
For details, see the text.}
\end{figure}

We now compare, in Fig.\ \ref{fig:sigCFNi}, the CF cross section of our calculations to the experimental data reported in Ref.\ \cite{MRR14}.
The full line depicts our CDCC results using $\varepsilon_{\rm max}=15$ MeV, which is adopted in all the calculations below.
The agreement with the experimental results is very good, with significant deviations observed only at the highest energies.
To assess the net effect of the low binding energy of $^6$Li on CF, we also show the BPM cross section for the potential $V_{PT}(R)$
(green dashed-line). Since this potential neglects the cluster structure of the projectile, its BPM cross section does not contain the barrier lowering effect.
Then comparing the BPM and the CF cross sections, one assesses the net effect of the low binding energy of the projectile. 
As observed  for other weakly bound systems~\cite{CGD15,HOM22,Mor18}, one finds enhancement at sub-barrier energies
and suppression above the Coulomb barrier. At the highest energies considered in Fig.~\ref{fig:sigCFNi}, one finds a suppression 
factor of 0.77.\\

 \begin{figure}
\includegraphics[width=8.5cm,angle=0]{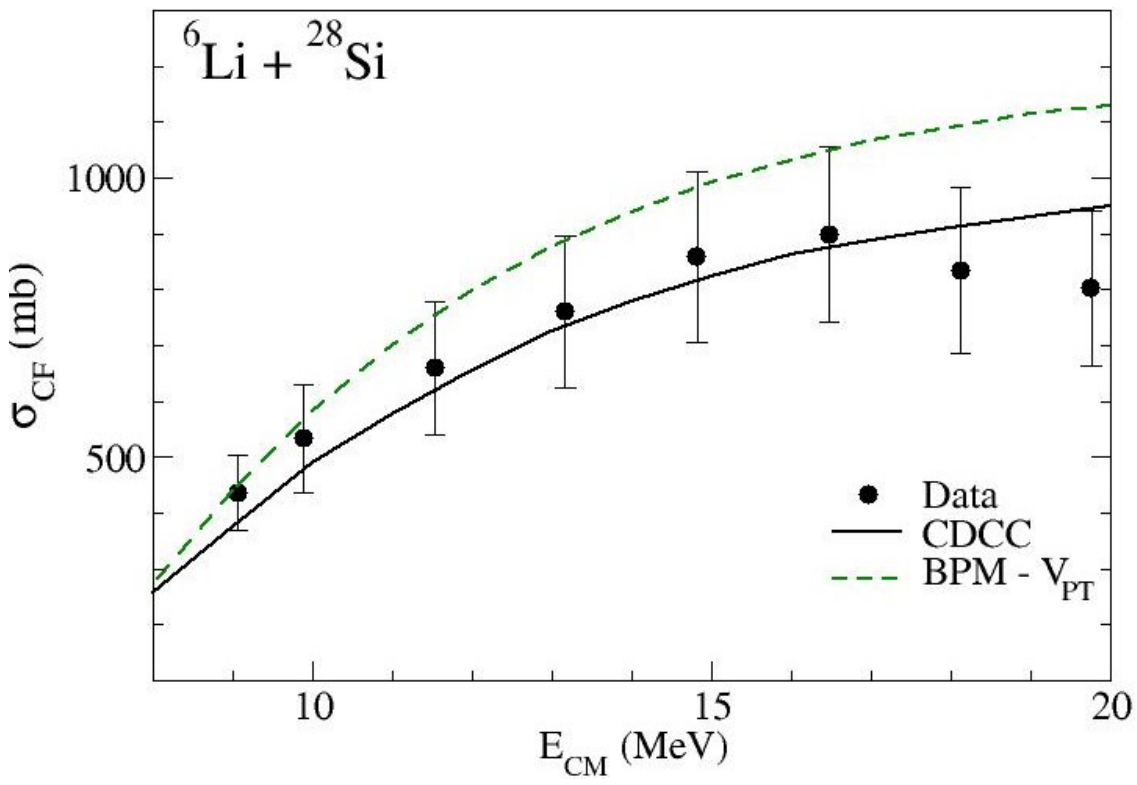}
\caption{\label{fig:sigCFSi} (Color online) Same as Fig.\ \ref{fig:sigCFNi} for the $^6$Li + $^{28}$Si Nuclei.
The data are from Ref.\ \cite{SiL17,SMB10}.
For details, see the text.}
\end{figure}

In order to check the validity of the above conclusions to collisions involving lighter targets, we have also carried out similar CDCC calculations for the $^6$Li + $^{28}$Si system.
Our findings concerning the convergence of the calculated cross-sections also hold in this case. Figure \ref{fig:sigCFSi} exhibits a comparison of the CF cross section of
our CDCC calculations to the CF data of Refs.\ \cite{SiL17,SMB10}. The data and the calculations 
 are for collision energies between 8 and 20 MeV, which are higher than the Coulomb barriers $V_{00}^B$ and  $V_{PT}^{B}$ for the $^6$Li + $^{28}$Si 
system (see Table~\ref{barriers}). 
The agreement between theory and experiment is very good, except at the highest collision energies, where the CDCC calculations overpredict the data. 
The figure also shows the cross section of the BPM with the potential $V_{PT}(R)$ (green dashed line). Comparing it to the CF cross sections 
we find that the latter is enhanced at sub-barrier energies and suppressed above the barrier, as in the collision with the $^{64}$Ni target. In this case, the
suppression factors at the highest energies is 0.84, which is slightly larger than for the heavier $^{64}$Ni target. This indicates that the suppression for the lighter
target is weaker. \\

Thus, our CDCC calculations give a fairly accurate description of the experimental CF cross-sections employing states in the continuum with energy up to 15 MeV.

\end{section}

\begin{section}{Concluding Remarks}
\label{sect:conclusions}

We have studied the CF cross sections in collisions of $^6$Li projectiles with $^{64}$Ni and $^{28}$Si targets. For this purpose, we performed CDCC calculations
using the pseudo states approach, and evaluated the CF cross sections using the CF-ICF model. 
We have investigated the influence of the largest energy $\varepsilon_{\rm max}$ allowed in the continuum states in CDCC calculations on the complete fusion cross-sections.
Collisions of $^6$Li projectiles with medium ($^{64}$Ni) and light ($^{28}$Si) targets have been considered.
In both cases, we found that the complete fusion is very sensitive to $\varepsilon_{\rm max}$ in the range $0 \le \varepsilon_{\rm max} \lesssim 5$ MeV.
Convergence is virtually reached between 10 MeV to 15 MeV, so that values around 10 MeV lead to a fairly good approximation to the converged results as the deviations are on the order of $5\%$.
Nevertheless, closed channels must be included in the calculations at low collision energies as their removal leads typically to 20\% deviations in the cases we examined.
The good agreement of the theoretical calculations with the available experimental data for the two systems studied gives support to the robustness of our conclusions.

\end{section}

\begin{acknowledgments}
This work was supported in part by the Brazilian agencies Conselho Nacional de Desenvolvimento Cient\'\i ­fico e Tecnol\'ogico (CNPq) ,  the Uruguayan agencies Programa de Desarrollo de las Ciencias B\'asicas (PEDECIBA) and the Agencia Nacional de Investigaci\'on e Innovaci\'on (ANII) for partial financial support.
This work has been done as a part of the project INCT-FNA, Proc. No.464898/2014-5.
We also thank the N\'ucleo Avan\c cado de Computa\c c\~ao de Alto Desempenho (NACAD), Instituto Alberto Luiz Coimbra de 
P\'os-Gradua\c c\~ao e Pesquisa em Engenharia (COPPE), Universidade Federal do Rio de Janeiro (UFRJ), for the use  of the supercomputer Lobo Carneiro and VERSATUS HPC where part of the calculations have been carried out.

\end{acknowledgments}

\bibliography{fusbreak-2023_vs04}
\bibliographystyle{apsrev4-2}

\end{document}